\RequirePackage{fix-cm}
\documentclass{svjour3}  
\usepackage[a4paper, total={6in, 8in}]{geometry} 
\smartqed  
\usepackage{graphicx}
\usepackage{dcolumn}
\usepackage{bm}
\usepackage{amsmath}

\usepackage{braket}
\usepackage[square,sort,comma,numbers]{natbib}

\def\bcen{\begin{center}}
\def\ecen{\end{center}}
\usepackage{epstopdf}
\begin{document}

\title{Simulation of single photon dynamics in coupled cavities through IBM quantum computer
}


\author{Nilakantha Meher \and
        Bikash K. Behera \and   
        Prasanta K. Panigrahi 
}


\institute{$^1$Department of Physics, Indian Institute of Technology Kanpur, Kanpur 208016, India,\\ $^2$Bikash's Quantum (OPC) Pvt. Ltd., Balindi, Mohanpur 741246, Nadia, West Bengal, India,\\$^3$Department of Physical Sciences,
Indian Institute of Science Education and Research Kolkata, Mohanpur 741246, West Bengal, India,\\ 
              \email{$^1$nilakantha.meher6@gmail.com}           
           \and
\email{$^2$bikash@bikashsquantum.com} \and \email{$^3$pprasanta@iiserkol.ac.in}
}

\date{Received: date / Accepted: date}

\maketitle

\begin{abstract}
We design a quantum circuit in IBM quantum computer that mimics the dynamics of single photon in a coupled cavity system. By suitably choosing the gate parameters in the quantum circuit, we could transfer an unknown qubit state between the qubits. The condition for perfect state transfer is obtained by solving the unitary time dynamics governed by the Hamiltonian of the coupled cavity system.  We then demonstrate the dynamics of entanglement between the two-qubits and show violation of Bell's inequality in IBM quantum computer.
\end{abstract}

\section{Introduction}
IBM has developed prototypes of 5-, 16-, 20- and 50-qubit quantum computers that are accessible through a cloud-based web-interface called IBM
 quantum experience \cite{IBMQE}.  
 This has attracted a great attention for verifying and testing quantum information protocols. Recently, many interesting protocols such as quantum repeater \cite{Behera2019}, Deutsch-Jozsa algorithm \cite{Gangopadhyay2018}, entanglement purification \cite{Behera2019}, quantum router \cite{Behera2019a}, error correction \cite{Anandu}, quantum teleportation \cite{Huang2020}, quantum cheque \cite{Behera2017}, quantum voting \cite{Joy2019}, etc. have been demonstrated using the IBM platform. Quantum computer has also been used for demonstrating many fundamental aspects of quantum mechanics, such as quantum tunneling \cite{Hegade}, simulating Klein-Gordon equation \cite{Kapil}, etc. Moreover, IBM quantum computer is shown to be useful for playing various games \cite{Mahanti2019,Paul2,Pal2020EPL}. 
 
In this article, we design a quantum circuit in IBM quantum computer that mimics the dynamics of single photon in a coupled cavity system. The circuit consists of various single-qubit and two-qubit quantum gates. By suitably choosing the gate parameters, we achieve perfect state transfer between the qubits. The condition for perfect state transfer is obtained by solving the unitary time dynamics of single photon in a coupled cavity system.
Here, the state that we consider to transfer from one qubit to the other qubit in the circuit is of the form $\alpha\ket{0}+\beta \ket{1}$. We characterize the success probability of state transfer in terms of fidelity between the target state and the experimentally obtained state (output from the circuit).  We then demonstrate the dynamics of entanglement between the two qubits and show the violation of Bell's inequality in IBM quantum computer.

The paper is organized as follows. In Sec. \ref{PhysicalModel}, we consider a system of coupled cavities and derive the condition for perfect transfer of an unknown state between the cavities, which will be used for transferring the same in the IBM platform. In Sec. \ref{Simulation}, we simulate the dynamics of single photon in a coupled cavity system through IBM quantum computer and realize quantum state transfer in Sec. \ref{State transfer}. Using the same circuit, we study the dynamics of entanglement and demonstrate violation of Bell's inequality in Sec. \ref{Entanglement}. Finally, we summarize our results and conclude with directions for future work in Sec. \ref{Summary}.
\section{Coupled cavity system}\label{PhysicalModel}
Here, we consider a system of coupled cavities, described by the Hamiltonian \cite{Ogd,Meher}
\begin{align}
H=\omega_1 a_1^\dagger a_1+\omega_2 a_2^\dagger a_2 +J(a_1^\dagger a_2+a_1 a_2^\dagger),
\end{align}
where $\omega_1 (\omega_2)$ is the resonance frequency of the first (second) cavity and $J$ is the coupling strength between the cavities. The operator $a_j (a_j^\dagger)$ is the annihilation (creation) operator, corresponding to the $j$th cavity. The operator $N=a_1^\dagger a_1+a_2^\dagger a_2$  commutes with the Hamiltonian $[H,N]=0$, hence, the operator $N$ is a constant of motion. As a consequence, the total number of photons remains constant during time evolution of the coupled cavity system. It is possible to find an invariant subspace having a fixed number of photons to diagonalize the Hamiltonian. Here, we restrict our attention up to single photon subspace and hence, the basis vectors are $\{\ket{0 0},\ket{1 0},\ket{0 1 }\}$. Here, $\ket{n  m }$ represents the first and second cavities containing $n$ and $m$ number of photons respectively. Now, we can write the Hamiltonian in the basis \{$\ket{0  0 },\ket{1  0 },\ket{0 1 }\}$ as
\begin{align}
H=\left[\begin{array}{ccc}
0 & 0 & 0\\
0 & \omega & J\\
0 & J & \omega
\end{array} \right],
\end{align}
where we assume $\omega_1=\omega_2=\omega $. The unitary matrix for time evolution takes the form
\begin{align}\label{Unitary}
U=e^{-iHt}=\left[\begin{array}{ccc}
1 & 0 & 0\\
0 & e^{-i\omega t}\cos Jt & -ie^{-i\omega t}\sin Jt \\
0 & -ie^{-i\omega t}\sin Jt & e^{-i\omega t}\cos Jt
\end{array} \right].
\end{align}

Consider the state of the first cavity to be of the form $\alpha \ket{0} +e^{i\eta}\beta \ket{1} $, and the second cavity in vacuum $\ket{0} $. Here, $\alpha$ and $\beta$ are two real numbers, satisfying $\alpha^2+\beta^2=1$.  So, the initial state is $\ket{\psi_{in}}=\alpha \ket{0}  \ket{0} +e^{i\eta}\beta \ket{1} \ket{0} $. Then, the evolved state at time $t$ is
\begin{align}\label{EvolvedState}
U\ket{\psi_{in}}&=\alpha \ket{0}  \ket{0} +e^{i\eta}\beta e^{-i\omega t}[\cos Jt \ket{1} \ket{0} -i\sin Jt \ket{0} \ket{1} ].
\end{align} 

Perfect state transfer is said to be realized if the evolved state becomes $\ket{\psi_{target}}=\alpha\ket{0} \ket{0} +e^{i\eta}\beta \ket{0} \ket{1} $, \textit{i.e.,} the state of the first cavity becomes a vacuum and the state of the second cavity becomes $\alpha \ket{0} +e^{i\eta}\beta \ket{1} $. This is possible if we choose $\omega/J=4k-1$ and $t=\pi/2J$. Hence, by properly choosing the system parameters such as resonance frequencies, coupling strength and evolution time, we can perfectly transfer an unknown quantum state from one cavity to the other cavity.


\section{Simulation in IBM quantum computer}\label{Simulation} 
\begin{figure}
\includegraphics[scale=0.38]{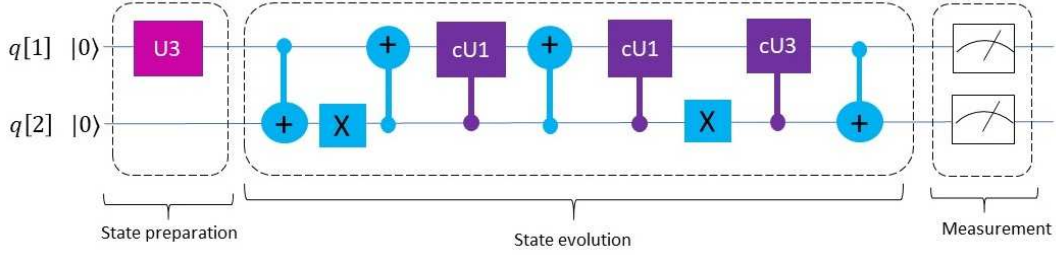}
\caption{Circuit diagram that mimics the dynamics of single photon in coupled cavities. The first part is for preparing a desired qubit state of the first qubit, and the second part does the quantum evolution of the two qubits, and the third part is the measurement of the two-qubits.}
\label{Circuitdiagram}
\end{figure}

The circuit that mimics the dynamics of single photon in a coupled cavity system is given in Fig. \ref{Circuitdiagram}. As, by default, the states of the qubits are $\ket{0}$ in the IBM quantum computer, we need to use suitable quantum gates to prepare a desired initial state. The first part in the circuit consists of a $U_3$ gate, which acts on the first qubit to prepare the initial state.  The second part of the circuit performs the quantum state evolution, and the last part is the measurement of the qubits. The circuit consists of various quantum gates, whose unitary matrices are of the form 
\begin{align}
&U_3(\theta',\phi',\lambda')=\left[\begin{array}{cc}
\cos \frac{\theta'}{2} & -e^{i\lambda'}\sin\frac{\theta'}{2}\\
e^{i\phi'}\sin\frac{\theta'}{2} & e^{i\phi'+i\lambda'}\cos\frac{\theta'}{2}
\end{array}\right],
X=\left[\begin{array}{cc}
0 & 1\\1 & 0
\end{array}\right], U_{CNOT}=\left[\begin{array}{cccc}
1 & 0 & 0 & 0\\0 & 1 & 0 & 0\\ 0 & 0 & 0 & 1\\ 0 & 0 & 1 & 0
\end{array}\right],\nonumber\\
& cU_1(\delta)=\left[\begin{array}{cccc}
1 & 0 & 0 & 0\\0 & 1 & 0 & 0\\ 0 & 0 & 1 & 0\\ 0 & 0 & 0 & e^{i\delta}
\end{array}\right], cU_3(\theta,\phi,\lambda)=\left[\begin{array}{cccc}
1 & 0 & 0 & 0\\0 & 1 & 0 & 0\\ 0 & 0 & \cos \frac{\theta}{2} & -e^{i\lambda}\sin \frac{\theta}{2}\\ 0 & 0 & e^{i\phi}\sin\frac{\theta}{2} & e^{i\phi+i\lambda}\cos \frac{\theta}{2}
\end{array}\right].
\end{align}
\noindent \textit{State preparation:} In IBM quantum computer, the initial state of the qubits are $\ket{0}$. Hence, we use a $U_3$ gate, which acts on the first qubit to prepare the state of the form $\alpha \ket{0}+e^{i\eta} \beta \ket{1}$, where $\alpha=\cos \theta'/2$, $\beta=\sin \theta'/2$ and $\eta=\phi'$. The initial combined state of both the qubits is  
\begin{align}
\ket{\psi_{in}}=\alpha\ket{00}+e^{i\eta}\beta\ket{10}.
\end{align}

\noindent \textit{State evolution:} Now, both the qubits pass through a series of quantum gates as can be seen in Fig. \ref{Circuitdiagram}. We calculate the output state of the circuit before the measurement is done,
\begin{align}\label{EvolvedStateExpt}
\ket{\psi_{out}}&=\alpha e^{i\delta} \ket{00}+e^{i\eta}\beta\left(e^{i\phi+i\lambda}\cos\theta/2\ket{10}-e^{i\lambda}\sin\theta/2\ket{01}\right),\nonumber\\
&=\alpha \ket{00}+e^{-i\delta}e^{i\eta}\beta\left(e^{i\phi+i\lambda}\cos\theta/2\ket{10}-e^{i\lambda}\sin\theta/2\ket{01}\right).
\end{align} 
Comparing this with Eqn. \ref{EvolvedState}, we can relate the gate parameters to the cavity parameters such that the circuit can produce the results of the coupled cavity system. This requires $\delta=\omega t, \theta=2Jt$, $\phi=-\pi/2$ and $\lambda=\pi/2$.

\begin{figure}
\centering 
\includegraphics[scale=0.42]{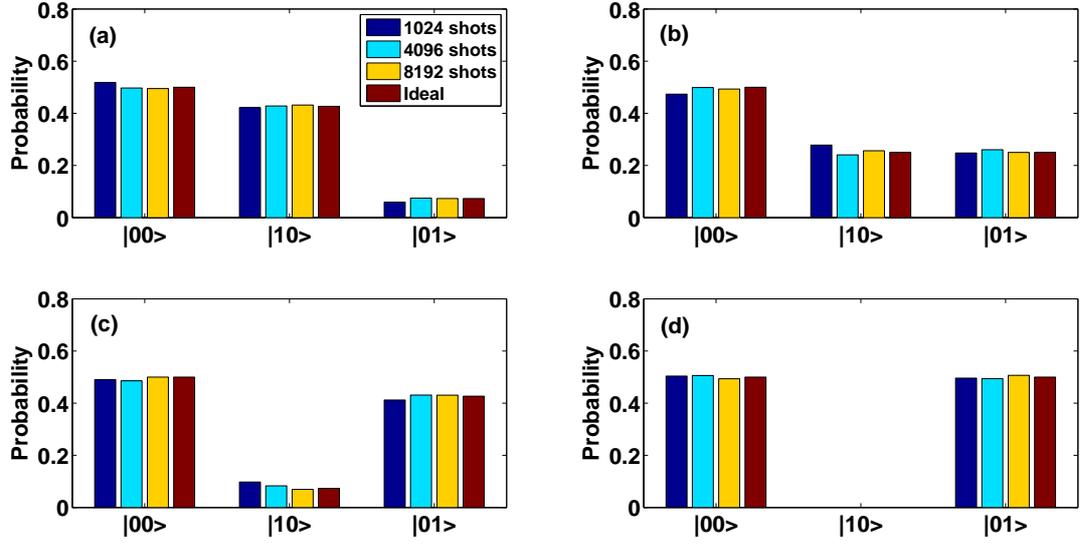}
\caption{Probabilities $P_{00},P_{10}$ and $P_{01}$ for various values of $\theta$ with different number of shots and compared with the ideal case. $(a) \theta=\pi/4$, $(b) \theta=\pi/2$, $(c)\theta=3\pi/4$ and $(d)\theta=\pi$. We choose $\theta'=\pi/2, \lambda'=0$ and $\phi'=0$ in $U_3$ gate, and $\phi=-\pi/2, \lambda=\pi/2$ in $cU_3$ gate and $\delta=\theta$ in $cU_1$ gate.  }
\label{ProbabilityVtheta}
\end{figure}

\noindent \textit{Measurement:} In IBM quantum computer, the measurement collapses the qubit state to either $\ket{0}$ and $\ket{1}$ with their corresponding probabilities.  From the evolved state given in Eqn. \ref{EvolvedStateExpt}, the probabilities of detecting various states of the two qubits such as $\ket{00},\ket{10}$ and $\ket{01}$ can be calculated as
\begin{align}
P_{00}=&|\bra{00}\psi(t)\rangle|^2=\alpha^2,\nonumber\\
P_{10}=&|\bra{10}\psi(t)\rangle|^2=\beta^2\cos^2\theta/2, \nonumber\\
P_{01}=&|\bra{01}\psi(t)\rangle|^2=\beta^2\sin^2\theta/2.\nonumber
\end{align} 
These are the theoretical results that will be true in the ideal case.
Fig. \ref{ProbabilityVtheta} shows the experimental values of $P_{00},P_{10}$ and $P_{01}$, obtained in IBM quantum computer for various values of $\theta$, and compared with their corresponding theoretically obtained values using the above equations.  We choose $\alpha=\beta=1/\sqrt{2}$ and $\eta=0$, which require $\theta'=\pi/2$, $\lambda'=0$ and $\phi'=0$ in the $U_3$ gate. IBM allows to run the experiment for 1, 1024, 4096 and 8192 times (shots). In 1 shot, it does not predict the correct results. Hence, we need to run the experiment for a large number of times to get the results as close to the theoretical results. As can be seen from the figure, the experimental results approach the theoretical results if the number of shots are large.  
\section{Quantum state transfer}\label{State transfer}
Now, we will discuss the scheme of quantum state transfer in IBM quantum computer. We would like to transfer the state $\frac{1}{\sqrt{2}}(\ket{0}+i\ket{1})$ from first qubit to the second qubit. To prepare this state, we choose $\theta'=\pi/2, \phi'=\pi/2$ and $\lambda'=0$ in the $U_3$ gate. Then, the evolved state before the measurement is
\begin{align}\label{StatetransferExpt}
\ket{\psi}=\frac{1}{\sqrt{2}}\ket{00}+\frac{i}{\sqrt{2}}e^{-i\delta}\left(e^{i\phi+i\lambda}\cos\theta/2\ket{10}-e^{-i\lambda}\sin\theta/2\ket{01}\right). 
\end{align}
For $\theta=\pi, \phi=-\pi/2$, $\lambda=\pi/2$ and $\delta=(4k-1)\pi/2=99\pi/2$ (we choose $k=25$), Eqn. \ref{StatetransferExpt} becomes $\ket{\psi}=(\ket{00}+i\ket{01})/\sqrt{2}$. This indicates that the state $\frac{1}{\sqrt{2}}(\ket{0}+i\ket{1})$ gets transferred from the first qubit to the second qubit. Theoretically, the fidelity of quantum state transfer is unity, however, this may not be true if we implement this in the IBM quantum computer. Hence, we calculate the success probability of quantum state transfer by using the fidelity \cite{Steffen2006Science} 
\begin{align}
F=\sqrt{\sqrt{\rho_{T}}\rho\sqrt{\rho_{T}}},
\end{align}
where $\rho_{T}$ is the theoretically obtained density matrix, and $\rho$ will be obtained experimentally via state tomography. 

\begin{figure}
\centering 
\includegraphics[scale=0.48]{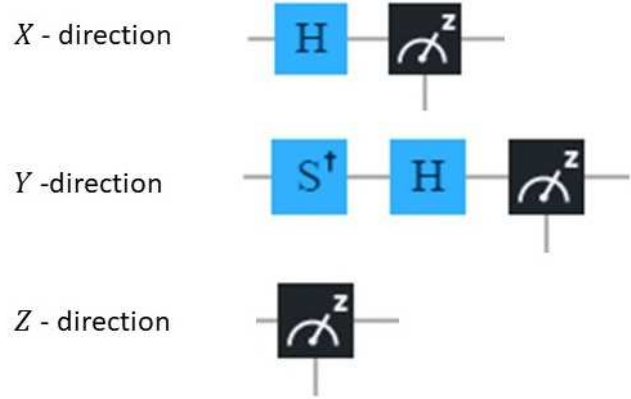}
\caption{Quantum gates required for measuring in different bases.}
\label{Basis}
\end{figure}

\begin{figure}
\centering 
\includegraphics[scale=0.4]{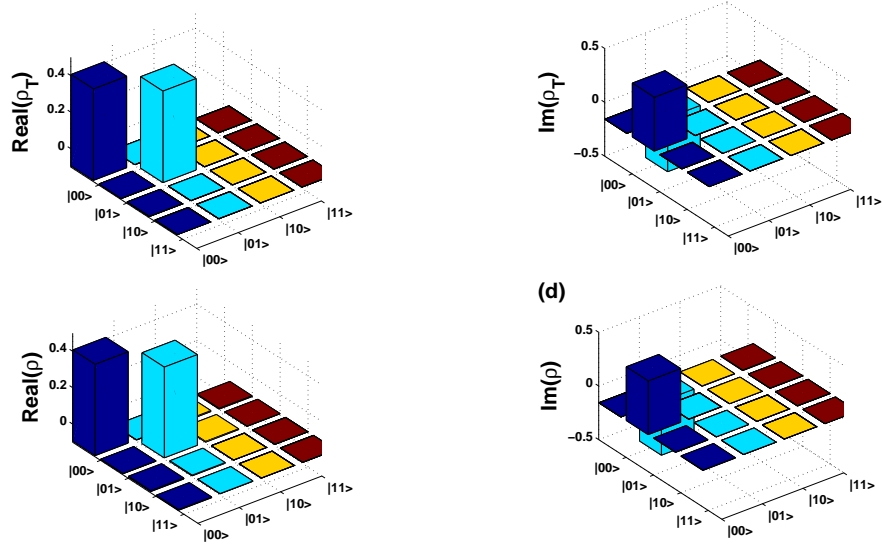}
\caption{The states of the two-qubit at $\theta=\pi$. We reconstruct the experimental density matrix by tomographic method. We choose $\delta=99\pi/2,\phi=-\pi/2$ and $\lambda=\pi/2$. $(a)$ and $(b)$ are the real and imaginary parts of the $\rho_T$. $(c)$ and $(d)$ are the real and imaginary parts of $\rho$.}
\label{Tomography}
\end{figure}

A general two-qubit density matrix can be written as \cite{James2001PRA}
\begin{align}
\rho&=\frac{1}{4}[I\otimes I+T_{IX}(I\otimes X)+T_{IY}(I\otimes Y)+T_{IZ}(I\otimes Z)+T_{XI}(X\otimes I)+T_{YI}(Y\otimes I)\nonumber\\
&~~~+T_{ZI}(Z\otimes I)+T_{XX}(X\otimes X)+T_{XY}(X\otimes Y)+T_{XZ}(X\otimes Z)+T_{YX}(Y\otimes X)\nonumber\\
&~~~+T_{YY}(Y\otimes Y)+T_{YZ}(Y\otimes Z)+T_{ZX}(Z\otimes X)+T_{ZY}(Z\otimes Y)+T_{ZZ}(Z\otimes Z)].
\end{align}
Here, $T_{IX},T_{IY},...$ are analogous to the Stokes parameter \cite{Andrew2007JOSAB,James2001PRA}.
It is to be noted that by measuring the parameters $T_{IX},T_{IY},...$, we can reconstruct the total density matrix $\rho$. But, measurement of these parameters requires to detect the state in various bases. For instance, to obtain $T_{IX}$, we need to measure only the second qubit in $X$-basis, while to obtain $T_{XY}$ we need to measure the first qubit in $X$-basis and the second qubit in $Y$-basis. Measuring in $X$-basis requires the use of a $H$ gate before the measurement, whereas measuring in $Y$-basis requires to use of $S^\dagger$ and $H$ gates (Refer Fig \ref{Basis}).
We can calculate the Stokes parameters using these probabilities \cite{Vishnu2018};
\begin{align}
T_{Ik}&=P_{0}-P_{1},\\
T_{jI}&=P_{0}-P_{1},\\
T_{jk}&=P_{00}-P_{01}-P_{10}+P_{11},
\end{align}     
where $j,k=X,Y,Z$. Using these parameters, we can reconstruct the total density matrix $\rho$. We compare the experimentally obtained density matrix with the theoretical density matrix in Fig. \ref{Tomography}. Here, $\rho_T=\ket{\psi_T}\bra{\psi_T}$, where $\psi_T=(\ket{00}+i\ket{01})/\sqrt{2}$. 
We calculate the fidelity between the theoretical density matrix and the experimentally obtained density matrix, and find it to be $F\approx 0.97$.

\section{Entanglement dynamics and violation of Bell's inequality}\label{Entanglement}
\begin{figure}
\centering 
\includegraphics[scale=0.4]{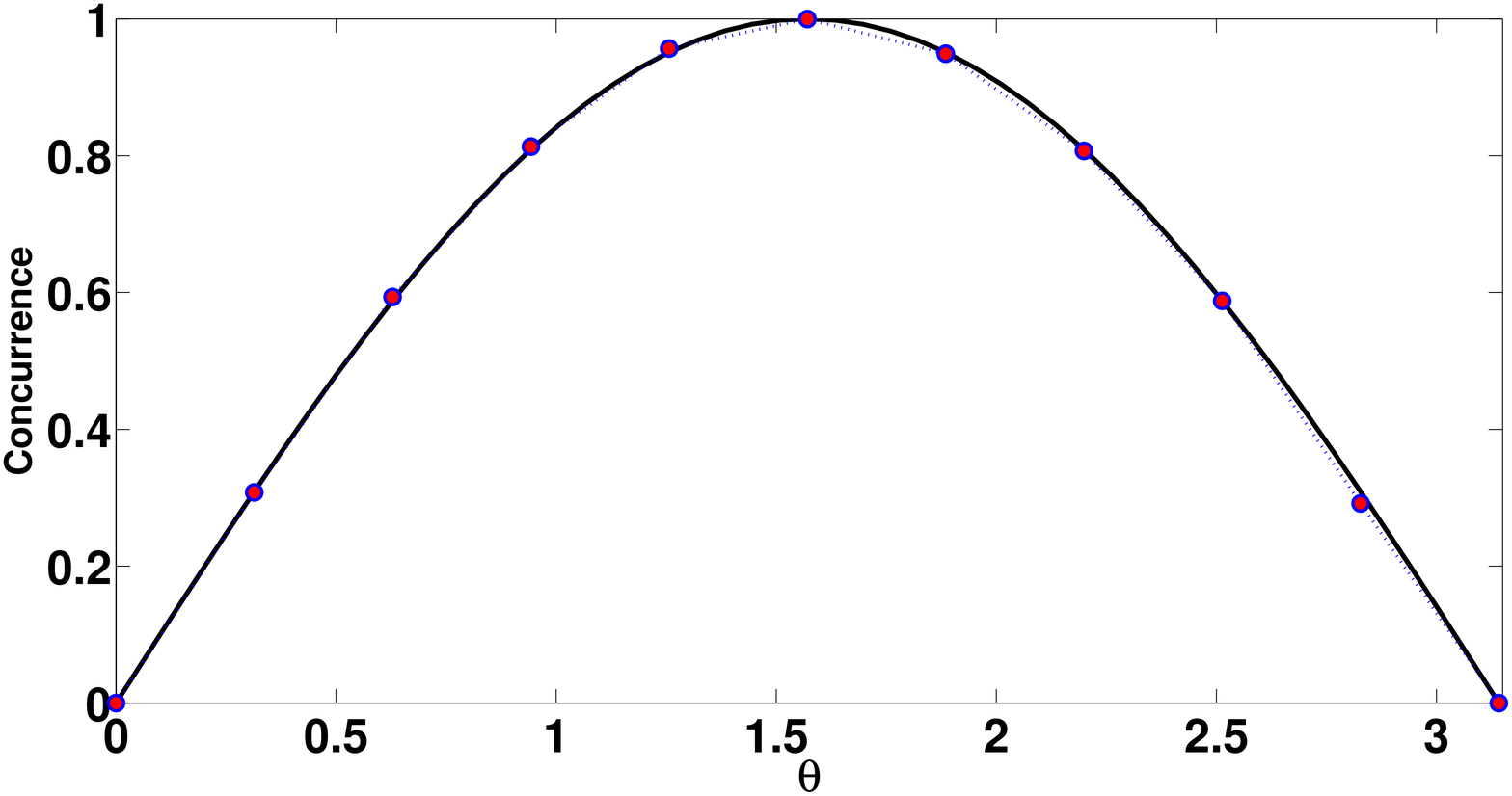}
\caption{Concurrence as a function of $\theta$ and compared with the expression $C=\sin\theta$. Concurrence is maximum at $\theta=\pi/2$.}
\label{ConcurrenceVstheta}
\end{figure}

In this section, we study the entanglement dynamics of the two qubits in the circuit given in Fig. \ref{Circuitdiagram}. We consider the initial state to be $\ket{10}$. Thus, we set $\theta'=\pi, \phi'=0$ and $\lambda'=0$ in $U_3$ gate. The state $\ket{10}$ passes through a series of gates. Then the output state before the measurement is
\begin{align}\label{EntState}
\ket{\psi_{out}}&=e^{i\phi+i\lambda}\cos\theta/2\ket{10}-e^{i\lambda}\sin\theta/2\ket{01}.
\end{align}
We choose $\phi=-\pi$ and $\lambda=\pi$.
The concurrence for the above state is \cite{Wootters1998PRL}
\begin{align}\label{ConcExp}
C=2\cos\theta/2 \sin\theta/2=2\sqrt{P_{10}}\sqrt{P_{01}},
\end{align}
where $P_{10}$ and $P_{01}$ are the probabilities of detecting the states $\ket{10}$ and $\ket{01}$. Hence, by measuring the probabilities of detecting the states $\ket{10}$ and $\ket{01}$, one can obtain the entanglement in the state. Fig. \ref{ConcurrenceVstheta} shows concurrence, measured experimentally as a function of $\theta$ and compared with the expression $C=\sin\theta$. As can be seen, the theoretical plot matches well with the experimentally obtained values. The state is maximally entangled at $\theta=\pi/2$.  
\begin{figure}[h!]
\centering 
\includegraphics[scale=0.4]{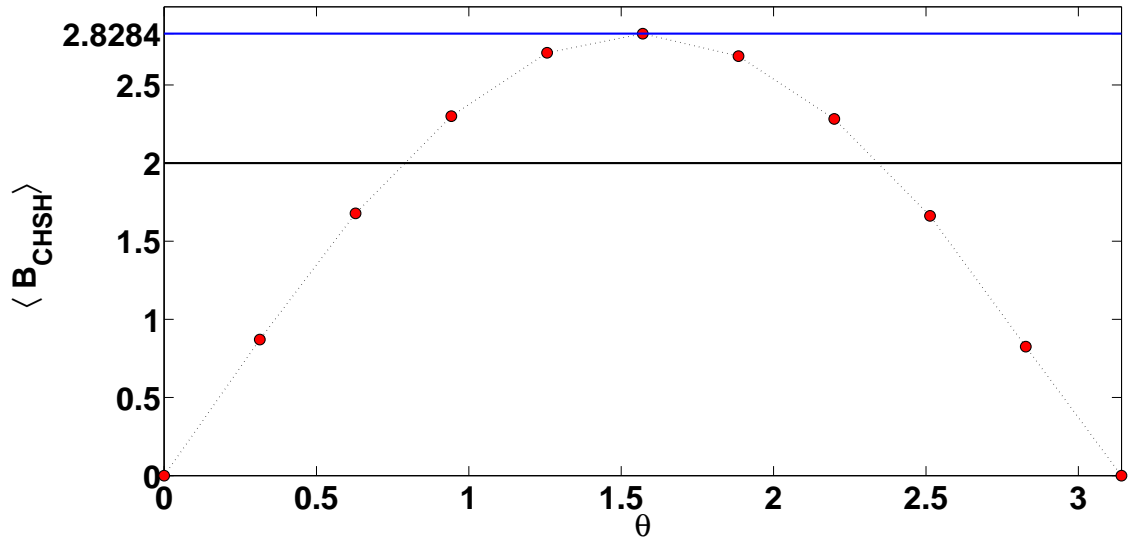}
\caption{Expectation value of the Bell parameter $\langle B_{CHSH}\rangle$ as a function of $\theta$. The upper line is $\langle B_{CHSH}\rangle=2\sqrt{2}$ and lower line is $\langle B_{CHSH}\rangle=2$. }
\label{BCHSHVstheta}
\end{figure}

The CHSH form of the Bell parameter is defined to be \cite{Bell,Clauser}
\begin{align}\label{BellCHSH}
B_{CHSH}=A_1 (B_1+B_2)+A_2(B_1-B_2),
\end{align}
where $A_i$ and $B_j$ are some observables correspond to the first and second qubits respectively.
For a local-realistic theory, $\langle B_{CHSH} \rangle \leq 2$. However, $\langle B_{CHSH} \rangle$ can be larger than 2 for entangled states \cite{Cirelson}. We take $A_1=\ket{0}\bra{1}+\ket{1}\bra{0}$, $A_2=i\ket{0}\bra{1}-i\ket{1}\bra{0}$, $B_1=\frac{1}{\sqrt{2}}[(1-i)\ket{0}\bra{1}+(1+i)\ket{1}\bra{0}]$ and $B_2=\frac{1}{\sqrt{2}}[(1+i)\ket{0}\bra{1}+(1-i)\ket{1}\bra{0}]$. The expectation  value of $B_{CHSH}$ in the state given in Eqn. \ref{EntState} is
\begin{align}
\langle B_{CHSH}\rangle=4\sqrt{2}\cos\theta/2 \sin\theta/2=4\sqrt{2}\sqrt{P_{10}}\sqrt{P_{01}}.
\end{align} 
Hence, by measuring $P_{01}$ and $P_{10}$, we can calculate the expectation value of the Bell parameter. The expectation value of the Bell parameter $\langle B_{CHSH}\rangle=4\sqrt{2}\sqrt{P_{10}}\sqrt{P_{01}}$ is shown as a function of $\theta$ in Fig. \ref{BCHSHVstheta}. Note that, $\langle B_{CHSH}\rangle$ reaches its maximum value $2\sqrt{2}$ for maximally entangled state $(\theta=\pi/2)$ \cite{Cirelson}. 

\section{Summary}\label{Summary}
We have successfully demonstrated the quantum state transfer protocol in IBM quantum computer. The suitable unitary operator for transferring the unknown qubit state is constructed by using the Hamiltonian of coupled cavities. Condition for perfect state transfer is obtained by solving the unitary time dynamics of the coupled cavities. By appropriately choosing the gate parameters, it is seen that the fidelity of quantum state transfer is of $0.97$.  We further studied the entanglement dynamics of the qubits in IBM quantum computer and demonstrated the violation of Bell's inequality. This scheme can be extended further to transfer an unknown qubit state between any two qubits in an array.


\begin{thebibliography}{10}
\providecommand{\url}[1]{{#1}}
\providecommand{\urlprefix}{URL }
\expandafter\ifx\csname urlstyle\endcsname\relax
  \providecommand{\doi}[1]{DOI~\discretionary{}{}{}#1}\else
  \providecommand{\doi}{DOI~\discretionary{}{}{}\begingroup
  \urlstyle{rm}\Url}\fi

\bibitem{IBMQE}
IBM Quantum Experience. http://research.ibm.com/ibm-q/

\bibitem{Anandu}
Anandu, K.M., Shaharukh, M., Behera, B.K., Panigrahi, P.K.: Demonstration of
  teleportation-based error correction in the ibm quantum computer.
\newblock arXiv \textbf{1902.01692} (2019)

\bibitem{Behera2017}
Behera, B.K., Banerjee, A., Panigrahi, P.K.: Experimental realization of
  quantum cheque using a five-qubit quantum computer.
\newblock Quantum Information Processing \textbf{16}(12), 312 (2017).
\newblock \urlprefix\url{https://doi.org/10.1007/s11128-017-1762-0}

\bibitem{Behera2019a}
Behera, B.K., Reza, T., Gupta, A., Panigrahi, P.K.: Designing quantum router in
  ibm quantum computer.
\newblock Quantum Information Processing \textbf{18}(11), 328 (2019).
\newblock \urlprefix\url{https://doi.org/10.1007/s11128-019-2436-x}

\bibitem{Behera2019}
Behera, B.K., Seth, S., Das, A., Panigrahi, P.K.: Demonstration of entanglement
  purification and swapping protocol to design quantum repeater in ibm quantum
  computer.
\newblock Quantum Information Processing \textbf{18}(4), 108 (2019).
\newblock \urlprefix\url{https://doi.org/10.1007/s11128-019-2229-2}

\bibitem{Bell}
Bell, J.S.: On the einstein podolsky rosen paradox.
\newblock Physics Physique Fizika \textbf{1}, 195--200 (1964).
\newblock \doi{10.1103/PhysicsPhysiqueFizika.1.195}.
\newblock
  \urlprefix\url{https://link.aps.org/doi/10.1103/PhysicsPhysiqueFizika.1.195}

\bibitem{Cirelson}
Cirel'son, B.S.: Quantum generalizations of bell's inequality.
\newblock Letters in Mathematical Physics \textbf{4}(2), 93--100 (1980).
\newblock \doi{10.1007/BF00417500}.
\newblock \urlprefix\url{https://doi.org/10.1007/BF00417500}

\bibitem{Clauser}
Clauser, J.F., Horne, M.A., Shimony, A., Holt, R.A.: Proposed experiment to
  test local hidden-variable theories.
\newblock Phys. Rev. Lett. \textbf{23}, 880--884 (1969).
\newblock \doi{10.1103/PhysRevLett.23.880}.
\newblock \urlprefix\url{https://link.aps.org/doi/10.1103/PhysRevLett.23.880}

\bibitem{Gangopadhyay2018}
Gangopadhyay, S., Manabputra, Behera, B.K., Panigrahi, P.K.: Generalization and
  demonstration of an entanglement-based deutsch-jozsa-like algorithm using a
  5-qubit quantum computer.
\newblock Quantum Information Processing \textbf{17}(7), 160 (2018).
\newblock \urlprefix\url{https://doi.org/10.1007/s11128-018-1932-8}

\bibitem{Hegade}
Hegade, N.N., Behera, B.K., Panigrahi, P.K.: Experimental demonstration of
  quantum tunneling in ibm quantum computer.
\newblock arXiv \textbf{1712.07326} (2017)

\bibitem{Huang2020}
Huang, N.N., Huang, W.H., Li, C.M.: Identification of networking quantum
  teleportation on 14-qubit ibm universal quantum computer.
\newblock Scientific Reports \textbf{10}(1), 3093 (2020).
\newblock \urlprefix\url{https://doi.org/10.1038/s41598-020-60061-y}

\bibitem{James2001PRA}
James, D.F.V., Kwiat, P.G., Munro, W.J., White, A.G.: Measurement of qubits.
\newblock Phys. Rev. A \textbf{64}, 052312 (2001).
\newblock \doi{10.1103/PhysRevA.64.052312}.
\newblock \urlprefix\url{https://link.aps.org/doi/10.1103/PhysRevA.64.052312}

\bibitem{Joy2019}
Joy, D., Sabir, M., Behera, B.K., Panigrahi, P.K.: Implementation of quantum
  secret sharing and quantum binary voting protocol in the ibm quantum
  computer.
\newblock Quantum Information Processing \textbf{19}(1), 33 (2019).
\newblock \urlprefix\url{https://doi.org/10.1007/s11128-019-2531-z}

\bibitem{Kapil}
Kapil, M., Behera, B.K., Panigrahi, P.K.: Quantum simulation of klein gordon
  equation and observation of klein paradox in ibm quantum computer.
\newblock arXiv \textbf{1807.00521} (2018)

\bibitem{Mahanti2019}
Mahanti, S., Das, S., Behera, B.K., Panigrahi, P.K.: Quantum robots can fly;
  play games: an ibm quantum experience.
\newblock Quantum Information Processing \textbf{18}(7), 219 (2019).
\newblock \urlprefix\url{https://doi.org/10.1007/s11128-019-2332-4}

\bibitem{Meher}
Meher, N., Sivakumar, S., Panigrahi, P.K.: Duality and quantum state
  engineering in cavity arrays.
\newblock Scientific Reports \textbf{7}(1), 9251-- (2017)

\bibitem{Ogd}
Ogden, C.D., Irish, E.K., Kim, M.S.: Dynamics in a coupled-cavity array.
\newblock Phys. Rev. A \textbf{78}, 063805 (2008)

\bibitem{Pal2020EPL}
Pal, A., Chandra, S., Mongia, V., Behera, B.K., Panigrahi, P.K.: Solving sudoku
  game using a hybrid classical-quantum algorithm.
\newblock Euro. Phys. Lett \textbf{128}(4), 40007 (2020)

\bibitem{Paul2}
Paul, S., Behera, B.K., Panigrahi, P.K.: Playing quantum monty hall game in a
  quantum computer.
\newblock arXiv \textbf{1901.01136} (2019)

\bibitem{Steffen2006Science}
Steffen, M., Ansmann, M., Bialczak, R.C., Katz, N., Lucero, E., McDermott, R.,
  Neeley, M., Weig, E.M., Cleland, A.N., Martinis, J.M.: Measurement of the
  entanglement of two superconducting qubits via state tomography.
\newblock Science \textbf{313}(5792), 1423--1425 (2006).
\newblock \doi{10.1126/science.1130886}.
\newblock \urlprefix\url{https://science.sciencemag.org/content/313/5792/1423}

\bibitem{Vishnu2018}
Vishnu, P.K., Joy, D., Behera, B.K., Panigrahi, P.K.: Experimental
  demonstration of non-local controlled-unitary quantum gates using a
  five-qubit quantum computer.
\newblock Quantum Information Processing \textbf{17}(10), 274 (2018).
\newblock \urlprefix\url{https://doi.org/10.1007/s11128-018-2051-2}

\bibitem{Andrew2007JOSAB}
White, A.G., Gilchrist, A., Pryde, G.J., O'Brien, J.L., Bremner, M.J.,
  Langford, N.K.: Measuring two-qubit gates.
\newblock J. Opt. Soc. Am. B \textbf{24}(2), 172--183 (2007).
\newblock \doi{10.1364/JOSAB.24.000172}.
\newblock \urlprefix\url{http://josab.osa.org/abstract.cfm?URI=josab-24-2-172}

\bibitem{Wootters1998PRL}
Wootters, W.K.: Entanglement of formation of an arbitrary state of two qubits.
\newblock Phys. Rev. Lett. \textbf{80}, 2245--2248 (1998).
\newblock \doi{10.1103/PhysRevLett.80.2245}.
\newblock \urlprefix\url{https://link.aps.org/doi/10.1103/PhysRevLett.80.2245}

\end{thebibliography}

\end{document}